\documentclass[journal]{tex/IEEEtran}  
\let\labelindent\relax
\usepackage{amsmath}
\usepackage{amssymb}
\usepackage{algorithm}
\usepackage{moresize}
\usepackage{tikz}
\usepackage{bm}
\usepackage{booktabs}
\usepackage[noend]{algpseudocode}
\usepackage{graphicx}
\usepackage{pgfplots}
\usepackage{pgfplotstable}
\usepackage[noadjust]{cite}
\usepgfplotslibrary{fillbetween} 
\usepackage[caption=false, font=footnotesize]{subfig}
\let\labelindent\relax
\usepackage[shortlabels]{enumitem}
\usepackage[colorinlistoftodos,prependcaption]{todonotes}
\IEEEoverridecommandlockouts                              

\algrenewcommand\textproc{\textsc}

\DeclareMathOperator*{\argmin}{arg\,min}

\newcommand{\myproof}{IEEEproof}

\title{Exact Worst-Case Execution-Time Analysis for Implicit Model Predictive Control}
\author{Daniel Arnstr\"om, David Broman, and Daniel Axehill
  \thanks{This work was partly supported by the Swedish Research Council (VR) under contract number 2017-04710.}
\thanks{D. Arnstr\"om and D. Axehill are with the Division of Automatic Control, Link\"oping University,
  Sweden 
{\tt\small daniel.\{arnstrom,axehill\}@liu.se}}%
\thanks{D. Broman is with the Division of Software and Computer Systems at KTH Royal Institute of Technology, Sweden
{\tt\small dbro@kth.se}}%
}

\begin{document}
\renewcommand{\baselinestretch}{1.0}

\definecolor{set19c1}{HTML}{E41A1C}
\definecolor{set19c2}{HTML}{377EB8}
\definecolor{set19c3}{HTML}{4DAF4A}
\definecolor{set19c4}{HTML}{984EA3}
\definecolor{set19c5}{HTML}{FF7F00}
\definecolor{set19c6}{HTML}{FFFF33}
\definecolor{set19c7}{HTML}{A65628}
\definecolor{set19c8}{HTML}{F781BF}
\definecolor{set19c9}{HTML}{999999}

\maketitle
\thispagestyle{empty}
\pagestyle{empty}
\newtheorem{proposition}{Proposition}
\newtheorem{lemma}{Lemma}
\newtheorem{corollary}{Corollary}
\newtheorem{remark}{Remark}
\newtheorem{theorem}{Theorem}
\newtheorem{definition}{Definition}
\newtheorem{assumption}{Assumption}
\newtheorem{example}{Example}

\begin{abstract}
    We propose the first method that determines the exact worst-case execution time (WCET) for implicit linear model predictive control (MPC). Such WCET bounds are imperative when MPC is used in real time to control safety-critical systems.  The proposed method applies when the quadratic programming solver in the MPC controller belongs to a family of well-established active-set solvers. For such solvers, we leverage a previously proposed complexity certification framework to generate a finite set of ``archetypal'' optimization problems; we prove that these archetypal problems form an \textit{execution-time equivalent cover} of all possible problems; that is, that they capture the execution time for solving any possible optimization problem that can be encountered online.
    Hence, by solving just these archetypal problems on the hardware on which the MPC is to be deployed, and by recording the execution times, we obtain the \textit{exact} WCET.
    In addition to providing formal proofs of the methods efficacy, we validate the method on an MPC example where an inverted pendulum on a cart is stabilized. The experiments highlight the following advantages compared with classical WCET methods: (i) in contrast to classical static methods, our method gives the \textit{exact} WCET; (ii) in contrast to classical measurement-based methods, our method guarantees a correct WCET estimate and requires fewer measurements on the hardware. 
\end{abstract}

\section{Introduction}
\label{sec:intro}
In Model Predictive Control (MPC) \cite{rawlings2017model}, an optimization problem is solved at each time step to produce a control action.
Over the last decades, advances in optimization algorithms and hardware implementations have enabled MPC to be applied in \textit{real time} on \textit{embedded} hardware, for example, in automotive and aerospace applications \cite{di2018real}. 
When the controlled system is, in addition to a real-time system, a \textit{safety-critical} system,  worst-case guarantees on the required time for solving the encountered optimization problems become imperative \cite{johansen2014dmpc}. 
Generally, the optimization problems are of the form   
\begin{equation}
    \label{eq:opt}
    \begin{aligned}
    \min_x f(x,\theta) \\
    g(x,\theta) \leq 0,
    \end{aligned}
\end{equation}
where the objective function $f: \mathbb{R}^{n_x} \times \mathbb{R}^{n_{\theta}} \to \mathbb{R}$ and the feasible set, defined by $g:\mathbb{R}^{n_x} \times \mathbb{R}^{n_{\theta}} \to \mathbb{R}^m$, depend on the decision variable $x$ (related to the control action) and the parameter $\theta$ (related to the state of the system being controlled). In other words, all possible optimization problems for a given MPC application are parametrized by $\theta \in \Theta_0 \subseteq \mathbb{R}^{n_{\theta}}$.
For $\emph{linear}$ MPC, which is the focus of this work, $f$ is quadratic and $g$ is affine, making \eqref{eq:opt} a \textit{quadratic program} (QP). 

To be able to solve problems of the form \eqref{eq:opt} in real time, several optimization strategies have been tailored for MPC applications. One such strategy, known as \emph{explicit} MPC \cite{bemporad2002explicit}, computes and stores the solution to \eqref{eq:opt} as a function of the parameter $\theta$. The online implementation then reduces to a lookup table, which trivially enables an exact worst-case complexity analysis. Explicit MPC is, however, restricted to small problems since the complexity of the solution grows quickly as the problems become larger. As a consequence, explicit MPC quickly exceeds the limited memory and computational resources available on embedded hardware \cite{cimini2017certqp}.

An alternative strategy, sometimes called \emph{implicit} MPC, embeds an optimization solver on the controlled system and solves problems of the form \eqref{eq:opt} at each time step for a given $\theta$. 
Solvers that have been specifically developed to solve QPs arising in MPC applications include, for example, \cite{domahidi2012efficient,ferreau2014qpoases,frison2020hpipm,patrinos2014accelerated,arnstrom2022daqp}.
Embedding an optimization solver circumvents the complexity explosion that explicit MPC suffers from, but also complicates worst-case analyses, since the behavior of the underlying optimization solver needs to be examined. The main focus of this work is on such worst-cases analyses for a popular class of methods used in implicit MPC: active-set methods \cite{nocedal,fletcher1971general,goldfarb1983dual,arnstrom2022daqp}. 

Traditionally, active-set methods have had the notorious drawback, especially unfavourable in real-time applications, that their worst-case computational complexity is exponential in the number of decision variables \cite{klee1972good}; the actual computational complexity observed in practice is, however, often far from this worst case \cite{spielman2004smoothed}. 
This theory/practice-gap was, in the context of linear MPC, closed in \cite{arnstrom2022unifying} by a framework that determines the \emph{exact} worst-case complexity. The framework includes, for example, the active-set methods in \cite{nocedal,fletcher1971general,goldfarb1983dual,arnstrom2022daqp}. 

While the framework in \cite{arnstrom2022unifying} can determine the exact number of iterations or FLOPs for solving any possible optimization problem for a given linear MPC application, these complexity measures do not directly translate to \textit{execution time}. For execution-time certificates, additional information such as the particular implementation of the optimization solver and the hardware on which it is implemented on need to be accounted for. 
To determine the worst-case execution time (WCET) of a program is known as \emph{the WCET problem} \cite{wilhelm2008worst},
and there are two main classes of methods to tackle it: \textit{static} methods and \textit{measurement-based} methods.

Static methods use program code and an abstract model of the hardware to determine possible executions paths of the program, which can then be used to determine the worst-case path in terms of execution time 
\cite{ferdinand2004ait,ballabriga2010otawa,broman2017brief}. The main advantage of static methods is that they provide guaranteed upper bounds on the worst-case execution time, given an accurate hardware model. Two drawbacks are that these worst-case bounds become more conservative when the program's complexity increases, and that developing accurate hardware models for a particular platform requires significant effort.

Measurement-based methods, on the other hand, execute the program from a particular set of inputs and processor states and records the execution time on the actual platform \cite{davis2019survey,natarajan2019weaklyhard}.
Two advantages of measurement-based methods are that they do not require an abstract model of the hardware, and that the measured execution times are exact for all the considered inputs and processor states. A major drawback is, however, that these methods cannot, generally, give guarantees on the worst-case execution time, since \textit{all possible} inputs and processor states cannot be explored within a reasonable time. 

\usetikzlibrary{shapes,arrows}
\tikzstyle{decision} = [diamond, draw, fill=white!20, 
text width=4.5em, text badly centered, node distance=3.5cm, inner sep=0pt]
\tikzstyle{block} = [rectangle, draw, fill=gray!10, 
text width=10em, text centered, rounded corners=0.2em, minimum height=2.5em,node distance=2.5cm]
\tikzstyle{object} = [rectangle, draw, fill=white!20, 
text width=10em, text centered, rounded corners=1em, minimum height=3em,node distance=2.5cm, inner sep=0mm]
\tikzstyle{ioi} = [trapezium, draw, trapezium right angle=120,rounded corners, node distance=2.5cm, text width=10em, minimum height=2.5em]

\tikzstyle{line} = [draw, -latex']

\begin{figure}
\begin{center}
\begin{tikzpicture}[scale=0.65, transform shape,node distance=3cm, auto]
    \node [object] (problem) at (-1,0) {Parametric problem of the form \eqref{eq:opt}};

    \node [block, above right of=problem, xshift=-4pt] (cert) {Certification method from \cite{arnstrom2022unifying}};
  \node [block, below right of=problem, xshift=-4pt] (codegen) {Code generation};

  \node [object, right of=cert, xshift=70pt] (probs) {Archetypal problems $\{\theta_i\}_{i=1}^N$};
  \node [object, right of=codegen, xshift=70pt] (program) {Program};
  
  \node [block, below right of=probs, xshift=0pt] (hw) {Measure execution time on hardware};

  \node[right of=hw,xshift=-5pt] (wcet) {WCET};

  \node[object,below right of=cert,xshift=6pt] (as-solver) {Active-set solver (Algorithm 1)};

  \path [line] (problem) -- (cert);
  \path [line] (problem) -- (codegen);
  \path [line] (as-solver) -- (cert);
  \path [line] (as-solver) -- (codegen);
  \path [line] (cert) -- (probs);
  \path [line] (codegen) -- (program);
  \path [line] (probs) -- (hw);
  \path [line] (program) -- (hw);
  \path [line] (hw) -- (wcet);
\end{tikzpicture}
\end{center}
\caption{Overview of the proposed method that determines the worst-case execution time (WCET) required to solve parametric optimization problems of the form \eqref{eq:opt} with any active-set solver covered by the framework in \cite{arnstrom2022unifying}.}
\label{fig:overview}
\end{figure}
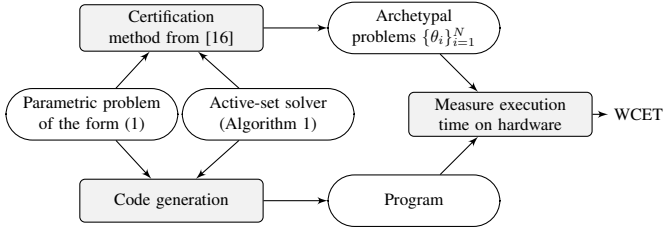

The main contribution of this work is a method that determines the \textit{exact} WCET for solving any problem instance in \eqref{eq:opt} when $f$ is quadratic and $g$ is affine. That the method determines the \textit{exact} WCET means that it is guaranteed to be sound without giving conservative results. The method use the certification framework in \cite{arnstrom2022unifying} to produce a finite set of archetypal problem instances of \eqref{eq:opt}. Given any active-set solver included in this framework, we prove that these archetypal problem instances capture \textit{all possible} execution paths and memory traces that the solver can generate (for \textit{any} possible problem instance); we say that these archetypal problem instances form an \textit{execution-time equivalent cover} of all possible problems. Hence, by solving just the archetypal problem instances on the hardware on which the solver is implemented, and by measuring the execution times, the \textit{exact} WCET can be determined, as visualized in Figure~\ref{fig:overview}. In contrast to static methods, our approach does not require an abstract hardware model, and does not yield a conservative WCET. In contrast to traditional measurement-based methods, our approach produces the \textit{exact} WCET.  From a practical point of view, the proposed method exceeds the method in \cite{arnstrom2022unifying} by generating bounds on the worst-case \textit{execution time}, instead of the higher level complexity measures, such as algorithm iterations and floating-point operations, considered therein.

Concretely, the contributions of this work are: 
\begin{enumerate}
    \item We introduce the concept of \textit{execution-time equivalent covers} for programs with parametrized inputs (Section~\ref{sec:wcet-problem}). 
    \item We show how the framework in \cite{arnstrom2022unifying} can be used to generate execution-time equivalent covers (Theorem \ref{th:main} and Corollary \ref{cor:wcet} in Section \ref{ssec:wcet-analysis}). 
        \item We present a concrete method (Algorithm~\ref{alg:proposed} in Section \ref{ssec:proposed-method}) that determines the \textit{exact} WCET of programs that realize any active-set method covered by \cite{arnstrom2022unifying}. 
        \item We propose a pruning technique to reduce the computational effort of Algorithm \ref{alg:proposed}, while still ensuring (through Lemma \ref{lem:subseq} in Section \ref{ssec:reduction}) a tight WCET estimate.  
        \item We apply Algorithm \ref{alg:proposed}  to determine the exact WCET of an MPC used to control an inverted pendulum on a cart (Section \ref{sec:result}); in particular, we use the dual-active set solver DAQP \cite{arnstrom2022daqp} in the MPC. 
\end{enumerate}

\section{The worst-case execution-time problem}
\label{sec:wcet-problem}
We define a \textit{program} as a realization of an algorithm, where we are, in particular, interested in algorithms that solve optimization problems of the form \eqref{eq:opt} for a given $\theta$. By ``a \textit{realization} of an algorithm'' we mean that a program, in addition to an underlying algorithm, depends on how the algorithm is implemented in code, and on the compiler/hardware used to produce/execute machine code.

For a given program $\mathcal{P}$ and a parametric optimization problem of the form \eqref{eq:opt}, the execution time for $\mathcal{P}$, starting in processor state $q$, required to solve the problem given by the parameter $\theta$ is denoted $\tau_{\mathcal{P}}(\theta,q)$. The WCET problem can then be formalized as solving the optimization problem 
\begin{equation}
    \label{eq:wcet}
    \bar{\tau}_{\mathcal{P}}  = \max_{\theta \in \Theta_0, q \in \mathcal{Q}} \tau_{\mathcal{P}}(\theta,q),
\end{equation}
where $\Theta_0$ parametrizes, as previously mentioned, all possible optimization problems that might need to be solved; and $\mathcal{Q}$ contains all possible starting processor states of the program. 
As \eqref{eq:wcet} highlights, there are two sources of uncertainty that complicate WCET analyses: ($i$) uncertainty in the input data, i.e, in the parameter $\theta$; ($ii$) uncertainty in the starting processor state, i.e., in $q$. 

\subsection{Uncertainty in input data}
In the context of MPC, the input data to a program is optimization problems of the form \eqref{eq:opt}, which are  parametrized by $\theta \in \Theta_0$. This parameter contains information such as the system state and setpoints, both of which exact values at an arbitrary time step are unknown; hence, any QP in \eqref{eq:opt} given by $\theta\in\Theta_0$ might need to be solved online, and the \textit{entirety} of $\Theta_0$ thus have to be accounted for in a WCET analysis.
Adding to the complexity, the system states and setpoints generally take values on a continuum, making it impossible to explore $\Theta_0$ with a finite number of samples. 
Despite this impossibility, we show that it is sufficient to only consider a finite subset of problems in $\Theta_0$ for solving the WCET problem in \eqref{eq:wcet} when considering programs that realize any active-set solver covered by \cite{arnstrom2022unifying}. For this we use the concept of execution-time equivalent problems: 
\begin{definition}[Exeuction-time equivalency]
Given a program $\mathcal{P}$ that solves \eqref{eq:opt}, two problem instances $\theta_1$, $\theta_2 \in \Theta_0$ are said to be \textit{execution-time equivalent} if $\tau_{\mathcal{P}}(\theta_1,q) = \tau_{\mathcal{P}}(\theta_2,q)$ for all processor states $q\in \mathcal{Q}$. 
A parameter set $\Theta \subseteq \Theta_0$ is execution-time equivalent if all parameters in $\Theta$ are pairwise execution-time equivalent. 
\end{definition}

In particular, we are interested in a finite set of problems such that $\mathcal{P}$ applied to this set of problems represents the behavior of $\mathcal{P}$ on the entire set $\Theta_0$. Such a set of problems can be found if an \textit{execution-time equivalent cover} of $\Theta_0$, defined below, is known.
\begin{definition}[Execution-time equivalent cover]
    Given a program $\mathcal{P}$ that solves \eqref{eq:opt}, a collection of regions $\{\Theta_i\}_{i=1}^N$ is said to be an execution-time equivalent cover of $\Theta_0$ if: $(i)$ $\cup_{i=1}^N \Theta_i = \Theta_0$; $(ii)$ $\Theta_i$ is execution-time equivalent.
\end{definition}

If there is an execution-time equivalent cover for a given program, finitely many problem instances are sufficient to determine the worst-case execution time for any problem in the possibly \textit{infinite} set of problems given by $\theta \in \Theta_0$, formalized in the following lemma:

\begin{lemma}
    \label{lem:etcover}
    Given an execution-time equivalent cover $\{\Theta_i\}_{i=1}^N$ of $\Theta_0$ to a program $\mathcal{P}$, and a known processor state $q$, the worst-case execution time $\bar{\tau}_{\mathcal{P}}(q) \triangleq \max_{\theta \in \Theta_0} \tau_{\mathcal{P}}(\theta,q)$ can be determined by querying $N$ problem instances. 
\end{lemma}
\begin{\myproof}
By selecting \textit{any} set of points $\{\theta_i\}_{i=1}^N$ such that $\theta_i \in \Theta_i$, we can by executing $\mathcal{P}$ for each $\theta_i$, with the starting processor state $q$, get the execution times $\{\tau_{\mathcal{P}}(\theta_i,q)\}_{i=1}^N$. This gives  
    \begin{equation*}
        \begin{aligned}
            \max_{i\in \{1,\dots,N\}} \tau_{\mathcal{P}}(\theta_i,q) 
            &= \max_{\theta \in \{\theta_i\}_{i=1}^N} \tau_{\mathcal{P}}(\theta,q)  
            = \max_{\theta \in  \cup_{i=1}^N \Theta_i} \tau_{\mathcal{P}}(\theta,q)\\
            &= \max_{\theta \in  \Theta_0} \tau_{\mathcal{P}}(\theta,q)
            = \bar{\tau}_{\mathcal{P}}(q),
        \end{aligned}
    \end{equation*}
    where we have used that each $\Theta_i$ is an execution-time equivalent set in the second equality, and that $\{\Theta_i\}_{i=1}^N$ is a cover of $\Theta_0$ in the third equality. 
\end{\myproof}

Unfortunately, finding an execution-time equivalent cover for a general program is difficult, often impossible, in practice. In Section~\ref{sec:wcet-as} we show, however, that it is possible to find such a cover when $\mathcal{P}$ is a realization of any active-set method covered by the framework in \cite{arnstrom2022unifying}, and when the optimization problem in \eqref{eq:opt} has an objective function $f$ that is quadratic and a constraint function $g$ that is affine (which is the case in linear MPC applications).

\subsection{Uncertainty in processor state}
Even if the uncertainty in the input data could be dealt with, uncertainty in the processor state also needs to be accounted for in the WCET problem \eqref{eq:wcet}. On modern processors, performance-enhancing features such as pipelining, caching, and multithreading, reduce execution predictability~\cite{axer2014building}, resulting in the execution time depending on previously executed instructions; in other words, the execution time is not history-invariant, which complicates WCET analyses. One approach to handle these uncertainties is to use predictable hardware architectures such as, e.g., FlexPRET \cite{zimmer2014flexpret}. In static methods, abstract interpretation \cite{cousot1977abstract}, is often used for giving sound but conservative estimations \cite{ferdinand1999efficient,touzeau2019cache}. 

In the proposed method, we deal with these uncertainties by starting in a deterministic processor state every time the program is called (i.e., every time an optimization problem of the form \eqref{eq:opt} is solved), formalized in the following assumption that is standard in the WCET field: 
\begin{assumption}[Deterministic starting state]
    \label{as:start-q-determ}
    There exists a known processor state $q_0$ that can be deterministically assigned online.
\end{assumption}

This assumption accounts for \textit{external} timing ambiguities between two calls to $\mathcal{P}$. Normally in static WCET methods, \textit{internal} timing ambiguities (for example from caching and pipelining) between basic blocks of the program also need to be accounted for explicitly. In the proposed method, we handle such internal timing ambiguities by determining the \textit{exact} sequence of processor instructions and the memory access patterns that are possible, given a deterministic starting state. The main contribution of this work (presented in Section~\ref{ssec:wcet-analysis}) is how a measurement-based approach together with the framework in \cite{arnstrom2022unifying} can determine such processor instruction sequences and the memory access patterns. 

The requirement of a deterministic processor state in Assumption~\ref{as:start-q-determ} can be relaxed by ensuring that the starting processor states used in the offline analysis are guaranteed to be ``worse'' than any possible starting processor state encountered online. This is formalized through the concept of dominance:  

\begin{definition}[Dominance]
    \label{def:dom}
For a program $\mathcal{P}$, a start state $q_1 \in \mathcal{Q}$ is said to \textit{dominate} another start state ${q_2 \in \mathcal{Q}}$ on $\Theta_0$ if ${\tau_{\mathcal{P}}(\theta,q_1) \geq  \tau_{\mathcal{P}}(\theta,q_2)}$ for all $\theta\in \Theta_0$.
\end{definition}

Accordingly, we formulate the following assumption as a relaxation of Assumption \ref{as:start-q-determ}:
\begin{assumption}[Dominated starting state]
    \label{as:start-q-dominated}
    There exists a known processor state $q_0$ that is dominated by all possible online starting processor states.
\end{assumption}

For simpler architectures, Assumption~\ref{as:start-q-dominated} can be ensured by flushing data and instruction caches, and emptying the pipeline before running the program (given that no \textit{timing anomalies} \cite{reineke2006definition} are possible.) 

Finally, we assume that no external factors, such as interrupts, are affecting the processor state while the program is solving a given problem instance. 
\begin{assumption}[Non-preemptive task]
    \label{as:interrupt}
    The program $\mathcal{P}$ is a \textit{non-preemptive task}. That is, it is never interrupted while running. 
\end{assumption}
\section{WCET Analysis for Active-Set Methods} 
\label{sec:wcet-as}
\subsection{Active-set methods}
\usetikzlibrary{shapes,arrows}
\tikzstyle{decision} = [diamond, draw, fill=white!20, 
text width=4.5em, text badly centered, node distance=2cm, inner sep=0pt]
\tikzstyle{block} = [rectangle, draw, fill=white!20, 
text width=10em, text centered, rounded corners, minimum height=4em]
\tikzstyle{decision} = [diamond, draw, fill=white!20, 
text width=4.5em, text badly centered, node distance=3.5cm, inner sep=0pt]
\tikzstyle{block} = [rectangle, draw, fill=white!20, 
text width=10em, text centered, rounded corners=0.2em, minimum height=2.5em,node distance=2.5cm]
\tikzstyle{start} = [rectangle, draw, fill=white!20, 
text width=10em, text centered, rounded corners=1em, minimum height=2.5em,node distance=2.5cm]
\tikzstyle{ioi} = [trapezium, draw, trapezium right angle=120,rounded corners, node distance=2.5cm, text width=10em, minimum height=2.5em]

\tikzstyle{line} = [draw, -latex']

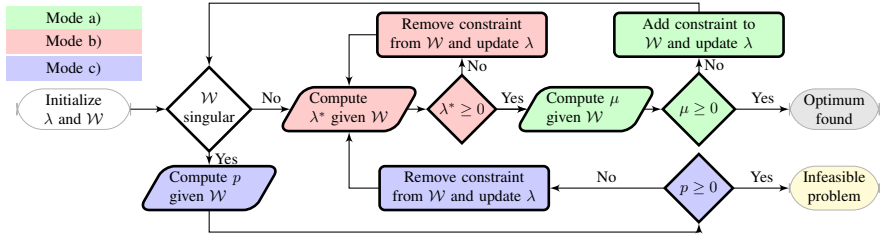
\begin{figure*}[htpb]
\begin{center}
\begin{tikzpicture}[scale=0.6, transform shape,node distance=5cm, auto]
  \node [start,line width=0mm,draw=gray!60,text width=6.5em] (init) at (-1,0) {Initialize $\lambda$ and $\mathcal{W}$};
  \node [decision, right of=init, line width=0.35mm, node distance=3cm,text width = 3.25em] (singular) {$\mathcal{W}$ singular };
  \node [ioi,fill=red!20,line width=0.35mm, right of=singular, node distance=3.1cm, text width=5em ] (add1) {Compute $\lambda^*$ given $\mathcal{W}$};
  \node [decision, right of=add1, fill=red!20,line width=0.35mm, node distance=2.5cm, text width=3.5em] (CSP) {$\lambda^* \geq 0$ };
  \node [ioi,fill=green!20,line width=0.35mm, right of=CSP, node distance=2.75cm, text width=5em] (lambda) {Compute $\mu$ given $\mathcal{W}$};
  \node [decision,fill=green!20,line width=0.35mm, right of=lambda, node distance=2.5cm, text width=3.75em] (dual) {$\mu \geq 0$};
  \node [block, line width=0.35mm, fill=green!20, above of=dual,node distance=1.65cm] (remove) {Add constraint to $\mathcal{W}$ and update $\lambda$};
  \node [start, fill=gray!20, right of=dual,line width=0mm,draw=gray!60,node distance=3cm, text width=5em] (end) {Optimum found};
  \node [block,fill=red!20,line width=0.35mm, above of=CSP,node distance=1.65cm] (add2) {Remove constraint from $\mathcal{W}$ and update $\lambda$};
  \node [right of=remove] (rm_help)  {};
  
  \node [ioi,fill=blue!20,line width=0.35mm, below of=singular,node distance=1.75cm, text width=5em] (sing-dir) {Compute $p$ given $\mathcal{W}$};
  \node [decision,fill=blue!20,line width=0.35mm, right of=sing-dir, node distance=10.85cm,text width=3.5em] (unb-check) {$p \geq 0$};
  \node [start,fill=yellow!20,line width=0mm, draw=gray!60, right of=unb-check,node distance=3cm, text width=5em] (unbounded) {Infeasible problem};
  \node [block, line width=0.35mm, fill=blue!20, left of=unb-check,node distance=5.2cm] (add-sing) {Remove constraint from $\mathcal{W}$ and update $\lambda$};
  \path [line] (init) -- (singular);
  \path [line] (singular) --node[above]{No} (add1);
  \path [line] (add1) -- (CSP);
  \path [line] (CSP) -- (lambda);
  \path [line] (lambda) -- (dual);
  \path [line] (dual) --node[above]{Yes} (end);
  \path [line] (dual) --node[right]{No} (remove);
  \path [line] (CSP) --node[above]{Yes} (lambda);
  \path [line] (CSP) --node[right]{No} (add2);
  \path [line] (add2) -| (add1); 
  \path [line] (remove) --++(0,0.7) -| (singular); 
  \path [line] (singular) --node[right]{Yes} (sing-dir); 
  \path [line] (sing-dir) --++(0,-0.975)-| (unb-check);
  \path [line] (unb-check) --node[above]{Yes} (unbounded);
  \path [line] (unb-check) --node[above]{No} (add-sing);
  \path [line] (add-sing) -| (add1);
  \node[rectangle,fill=green!20,minimum width=30mm] (aff) at (-1,2){Mode a)}; 
  \node[rectangle,fill=red!20,minimum width=30mm] (quad) at (-1,1.5){Mode b)}; 
  \node[rectangle,fill=blue!20,minimum width=30mm] (quad) at (-1,0.9){Mode c)}; 
\end{tikzpicture}
\end{center}
\caption{Flow chart characterizing Algorithm \ref{alg:prot-as} that gives a prototypical description of dual active-set method (e.g., \cite{goldfarb1983dual,arnstrom2022daqp}.)}
\label{fig:main-alg}
\end{figure*}

Active-set solvers, such as \cite{ferreau2014qpoases,nocedal,fletcher1971general,goldfarb1983dual,arnstrom2022daqp},  are commonly used in real-time MPC for solving problems of the form \eqref{eq:opt} for a given $\theta$. 
A prototypical description of \textit{dual} active-set methods that are covered by the certification method in \cite{arnstrom2022unifying} is given in Algorithm~\ref{alg:prot-as}, and its corresponding flow chart is shown in Figure~\ref{fig:main-alg}. The main idea behind active-set methods is to find constraints that hold with equality at the optimum, since this information is sufficient for finding a solution (see, e.g., \cite[Lemma 3.1]{arnstrom2021lic}). To this end, active-set methods update a so-called \emph{working set}, denoted $\mathcal{W}$, which contains indices of the constraints in \eqref{eq:opt} that are enforced to hold with equality (such constraints are said to be \emph{active}, hence the name \textit{active}-set methods.) In each iteration, a constraint is either added or removed to/from $\mathcal{W}$, until an optimizer is found. For details on active-set methods, see, for example, \cite{wong2011active} or \cite[Sec. 16.5]{nocedal}. Specifics regarding the dual active-set solver that we use in the experiments in Section~\ref{sec:result} can be found in \cite{arnstrom2022daqp}, in which the steps of Algorithm \ref{alg:prot-as} are concretized. 
\begin{algorithm}
  \caption{Prototypical dual active-set method}
  \label{alg:prot-as}
  \begin{algorithmic}[1]
      \Require $\lambda_0, \mathcal{W}_0$, problem of the form \eqref{eq:opt}, $\theta$  
    \Ensure $\lambda^*, \mathcal{W}^*$ 
    \State $k \leftarrow 0$
    \Repeat
    \If {$\mathcal{W}_k$ not singular}\label{step:sing-if}
  	    \algrenewcommand{\alglinenumber}[1]{\color{red}\footnotesize#1:}
        \State $\lambda^* \leftarrow$ \text{Solve lin. eq. system defined by $\mathcal{W}_k$} \label{step:linsys} 
        \If {$\lambda^* \geq 0$} \label{step:lam-if}
  	        \algrenewcommand{\alglinenumber}[1]{\color{green}\footnotesize#1:}
            \State $\mu \leftarrow$ Affine transform of $\lambda^*$ defined by $\mathcal{W}_k$ \label{step:affine-transformation} 
            \If {$\mu \geq 0$}\label{step:mu-if}
                \State \Return optimum found 
            \Else\:{($\lambda^* \ngeq 0$)}
            \State Add constraint to $\mathcal{W}$ and update $\lambda$.\label{step:add-constr} 
            \EndIf
  	    \algrenewcommand{\alglinenumber}[1]{\color{black}\footnotesize#1:}
        \Else
  	        \algrenewcommand{\alglinenumber}[1]{\color{red}\footnotesize#1:}
            \State Remove constraint from $\mathcal{W}$ and update $\lambda$. 
  	        \algrenewcommand{\alglinenumber}[1]{\color{black}\footnotesize#1:}
        \EndIf
    \Else{\:($\mathcal{W}_k$ singular)}
  	    \algrenewcommand{\alglinenumber}[1]{\color{blue}\footnotesize#1:}
        \State $p \leftarrow $ Solve lin. eq. system defined by $\mathcal{W}_k$ \label{step:linsys-sing}
        \If{$p \geq 0$} \Return infeasible problem \label{step:p-if}
        \EndIf
        \State Remove constraint from $\mathcal{W}$ and update $\lambda$.\label{step:rm-constr} 
  	    \algrenewcommand{\alglinenumber}[1]{\color{black}\footnotesize#1:}
    \EndIf
    \State $k\leftarrow k+1$
    \Until{termination}
  \end{algorithmic}
\end{algorithm}
\begin{remark}[Primal vs dual methods]
    Algorithm \ref{alg:prot-as} shows a prototypical \emph{dual}, rather than \emph{primal}, active-set method. Dual active-set methods work on a dual problem to \eqref{eq:opt}, which is of the form  
    $\min_{\lambda \geq 0} d(\lambda,\theta)$ (where $d$ is quadratic if $f$ is quadratic and $g$ is affine). The solution $x^*$ to \eqref{eq:opt} can be directly retrieved from the dual solution $\lambda^* \triangleq \argmin_{\lambda \geq 0} d(\lambda,\theta)$ (see, e.g., \cite[Eq. (2a)]{arnstrom2022daqp}). An advantage with the dual formulation is that the constraints are simple nonnegative constraints $\lambda \geq 0$ while the constraints in \eqref{eq:opt}, $g(x,\theta) \leq 0$, can be more involved. 
    The main results of this work do, however, hold for any active-set method that is covered by the framework in \cite{arnstrom2022unifying}, which include primal active-set methods.
\end{remark}
\begin{remark}[Singular working set]
    By a ``singular'' working set $\mathcal{W}$ at Step \ref{step:sing-if} in Algorithm \ref{alg:prot-as}, we mean that the corresponding \textit{reduced Hessian} (see Theorem 16.2 in \cite{nocedal}) is singular.
\end{remark}
\usetikzlibrary{shapes,arrows}
\tikzstyle{decision} = [diamond, draw, fill=white!20, 
text width=3.25em, text badly centered, node distance=2cm, inner sep=0pt, line width=0.35mm]
\tikzstyle{block} = [rectangle, draw, fill=white!20, 
text width=3em, text centered, rounded corners=0.2em, minimum height=2.5em,node distance=2cm,
line width=0.35mm]
\tikzstyle{start} = [rectangle, draw, fill=white!20, 
text width=3em, text centered, rounded corners=0.4em, minimum height=2.5em,node distance=2cm]
\tikzstyle{ioi} = [trapezium, draw, trapezium right angle=120,rounded corners, fill=blue!60, node distance=2cm, minimum height=2.7em] 
\tikzstyle{ioi} = [trapezium, draw, trapezium right angle=120, node distance=2cm, minimum height=2.5em, line width=0.35mm]
\begin{figure}[htpb]
\begin{center}
\begin{tikzpicture}[scale=0.35, transform shape,node distance=3.5cm, auto]
  \node [] (region) at (-3.5,0) {\HUGE $\theta \in \Theta_1$:};
  \node [start,line width=0mm,draw=gray!60] (init) at (-1,0) {};

  \node [decision, right of=init] (singular1) {};
  \node [ioi,fill=red!20,line width=0.35mm, right of=singular1] (b-linsys-1) {};
  \node [decision, right of=b-linsys-1, fill=red!20] (b-comp-1) {};

  \node [ioi,fill=green!20, right of=b-comp-1] (a-linsys-1) {};
  \node [decision,fill=green!20, right of=a-linsys-1] (a-comp-1) {};
  \node [block,fill=green!20, right of=a-comp-1] (a-change-1) {};
  
  \node [decision, right of=a-change-1] (singular2) {};
  \node [right of=singular2,node distance=2cm] (dots) {\Huge $\cdots$};
  \node [decision,fill=green!20, right of=dots] (a-comp-2) {};

  \node [start,fill=gray!20, right of=a-comp-2] (final) {};

  \path [line] (init) -- (singular1);
  \path [line] (singular1) -- (b-linsys-1);
  \path [line] (b-linsys-1) -- (b-comp-1);
  \path [line] (b-comp-1) -- (a-linsys-1);
  \path [line] (a-linsys-1) -- (a-comp-1);
  \path [line] (a-comp-1) -- (a-change-1);
  \path [line] (a-change-1) -- (singular2);
  \path [line] (singular2) -- (dots);
  \path [line] (dots) -- (a-comp-2);
  \path [line] (a-comp-2) -- (final);

  \node [] (region) at (-3.5,-2) {\HUGE $\theta \in \Theta_2$:};
  \node [start,line width=0mm,draw=gray!60] (init) at (-1,-2) {};

  \node [decision, right of=init] (singular1) {};
  \node [ioi,fill=blue!20,line width=0.35mm, right of=singular1] (b-linsys-1) {};
  \node [decision, right of=b-linsys-1, fill=blue!20] (b-comp-1) {};
  \node [block,fill=blue!20,line width=0.35mm, right of=b-comp-1] (c-change-1) {};

  \node [ioi,fill=red!20, right of=c-change-1] (a-linsys-1) {};
  \node [decision,fill=red!20, right of=a-linsys-1] (a-comp-1) {};
  \node [ioi,fill=green!20, right of=a-comp-1] (a-change-1) {};
  
  \node [right of=a-change-1,node distance=2cm] (dots) {\Huge $\cdots$};
  \node [decision,fill=green!20, right of=dots] (a-comp-2) {};

  \node [start,fill=gray!20, right of=a-comp-2] (final) {};

  \path [line] (init) -- (singular1);
  \path [line] (singular1) -- (b-linsys-1);
  \path [line] (b-linsys-1) -- (b-comp-1);
  \path [line] (b-comp-1) -- (c-change-1);
  \path [line] (c-change-1) -- (a-linsys-1);
  \path [line] (a-linsys-1) -- (a-comp-1);
  \path [line] (a-comp-1) -- (a-change-1);
  \path [line] (a-change-1) -- (dots);
  \path [line] (dots) -- (a-comp-2);
  \path [line] (a-comp-2) -- (final);
  
  \node [] (vdots1) at (-3,-3.25) {\HUGE $\vdots$};
  \node [] (vdots2) at (2,-3.25) {\HUGE $\vdots$};
  \node [] (vdots3) at (7,-3.25) {\HUGE $\vdots$};
  \node [] (vdots4) at (12,-3.25) {\HUGE $\vdots$};
  \node [] (vdots5) at (17,-3.25) {\HUGE $\vdots$};

  \node [] (region) at (-3.5,-4.5) {\HUGE $\theta \in \Theta_N$:};
  \node [start,line width=0mm,draw=gray!60] (init) at (-1,-4.5) {};

  \node [decision, right of=init] (singular1) {};
  \node [ioi,fill=red!20,line width=0.35mm, right of=singular1] (b-linsys-1) {};
  \node [decision, right of=b-linsys-1, fill=red!20] (b-comp-1) {};

  \node [ioi,fill=green!20, right of=b-comp-1] (a-linsys-1) {};
  \node [decision,fill=green!20, right of=a-linsys-1] (a-comp-1) {};
  \node [block,fill=green!20, right of=a-comp-1] (a-change-1) {};
  
  \node [decision, right of=a-change-1] (singular2) {};
  \node [right of=singular2,node distance=2cm] (dots) {\Huge $\cdots$};
  \node [decision,fill=blue!20, right of=dots] (a-comp-2) {};

  \node [start,fill=yellow!20, right of=a-comp-2] (final) {};

  \path [line] (init) -- (singular1);
  \path [line] (singular1) -- (b-linsys-1);
  \path [line] (b-linsys-1) -- (b-comp-1);
  \path [line] (b-comp-1) -- (a-linsys-1);
  \path [line] (a-linsys-1) -- (a-comp-1);
  \path [line] (a-comp-1) -- (a-change-1);
  \path [line] (a-change-1) -- (singular2);
  \path [line] (singular2) -- (dots);
  \path [line] (dots) -- (a-comp-2);
  \path [line] (a-comp-2) -- (final);
\end{tikzpicture}
\end{center}
\caption{Control flow of Algorithm \ref{alg:prot-as} for different optimization problems (parametrized by $\theta$) determined by \cite{arnstrom2022unifying}. The color and shape of a block corresponds to the colors and shapes of the blocks in the flow chart in Figure~\ref{fig:main-alg}.}
\label{fig:control-flows}
\end{figure}
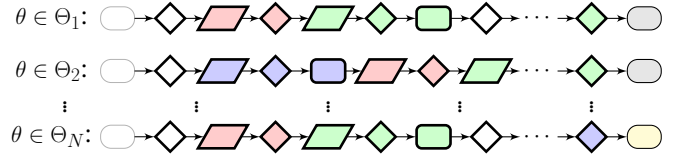

\subsection{Parametric complexity certification of active-set methods}
Since the QPs generated by \eqref{eq:opt} vary for different parameters ${\theta \in \Theta_0}$, Algorithm~\ref{alg:prot-as} will ``behave'' differently for different values of $\theta$; the certification method in \cite{arnstrom2022unifying} determines this parametric ``behavior''. To be more precise, the method determines the number of iterations $\bar{k}(\theta)$ and the working-set sequence $\{\mathcal{W}_i(\theta)\}_{i=0}^{\bar{k}(\theta)}$ for any $\theta \in \Theta_0$. 
In turn, the  working set sequence $\{\mathcal{W}_i(\theta)\}_{i=0}^{\bar{k}(\theta)}$ reveals the following information about Algorithm \ref{alg:prot-as} for all $\theta \in \Theta_0$: 
(i) The exact number of iterations $\bar{k}(\theta)$ in the outer loop; (ii) The decision made in the if-statements in Steps \ref{step:sing-if}, \ref{step:lam-if}, \ref{step:mu-if}, and \ref{step:p-if}; (iii) The dimensions of the linear systems solved in Step \ref{step:linsys} and \ref{step:linsys-sing} and the dimension of the affine transformations performed in Step \ref{step:affine-transformation}.

Put in a graphical way, the method in \cite{arnstrom2022unifying} determines \emph{exactly} the possible paths in the flow chart in Figure \ref{fig:main-alg}, from initialization until termination, when Algorithm \ref{alg:prot-as} is applied to \emph{any} problem given by $\theta \in \Theta_0$.
Not only is the possible paths provided, but also parameter regions $\{\Theta_i\}_{i=1}^N$ that represent optimization problems that generate the \emph{same} paths, visualized in Figure \ref{fig:control-flows}. 
The following proposition formalizes what the method in \cite{arnstrom2022unifying} outputs:

\begin{proposition}[Output from \cite{arnstrom2022unifying}]
    \label{prop:unify}
    The certification method in \cite{arnstrom2022unifying} outputs a collection of tuples 
    $\{(\bar{k}_i, \mathbb{W}_i, \Theta_i)\}_{i=1}^N$, where $\mathbb{W}_i$ is the working-set sequence $\{\mathcal{W}_i\}_{i=0}^{\bar{k}_i}$ generated by Algorithm \ref{alg:prot-as} for all $\theta \in \Theta_i$. 
    Moreover, the regions $\{\Theta_i\}_{i=1}^N$ form a partition of the nominal region $\Theta_0$ of interest. That is, $\cup_{i=1}^N \Theta_i = \Theta_0$ and $\mathring{\Theta}_i \cap \mathring{\Theta}_j=\emptyset$ when $i\neq j$ (where $\mathring \Theta$ denotes the interior of $\Theta$.) 
\end{proposition}

Next, we describe how the properties of the partition generated by \cite{arnstrom2022unifying} can be used to determine the worst-case \emph{execution time}; this is the main contribution of this work. 

\subsection{WCET analysis}
\label{ssec:wcet-analysis}
When moving from the iteration certificates derived in \cite{arnstrom2022unifying} to execution-time certificates, we need to move from an algorithmic perspective to a programmatic perspective. That is, we need to account for how the algorithm is implemented in code, how this code is compiled, and on which hardware the compiled machine code is executed. 

Instead of explicitly listing necessary requirements on the code implementation, compiler, and hardware for a program that realize Algorithm~\ref{alg:prot-as}, we use an implicit requirement: the processor instructions generated by the program should not depend \emph{directly} on the value of $\theta$, but only implicitly through the working-set sequence $\{\mathcal{W}_i(\theta)\}_{i=0}^{\bar{k}(\theta)}$. 
Another way of putting this is that the processor operations executed in any block in Figure~\ref{fig:main-alg} should not depend directly on the value of $\theta$: 
different values of $\theta$ will only change the values that are stored in memory/registers, but not the executed operations nor which memory addresses are accessed.
We formalize this requirement in the following assumption, and 
in Appendix \ref{ap:assumption-details} we expand upon how it can be ensured in practice. 

\begin{assumption}
    \label{as:p-as}
There exists a program $\mathcal{P}$ that realizes Algorithm~\ref{alg:prot-as} such that, given a particular processor state at the start of an iteration, the working-set $\mathcal{W}$ completely determines the sequence of executed processor instructions and memory access pattern in that iteration.
\end{assumption}

We are now ready to state the main result of this work, which allows us to go from iteration certificates to execution-time certificates:
\begin{theorem}
    \label{th:main}
    Assume that a program $\mathcal{P}$ satisfies Assumption~\ref{as:p-as}; then the partition $\{\Theta_i\}^N_{i=1}$ generated by the certification method in \cite{arnstrom2022unifying} is an execution-time equivalent cover of $\Theta_0$ for $\mathcal{P}$.
\end{theorem}
\begin{\myproof}
    Let $q_0$ denote the starting processor state for the program. Since all $\theta \in \Theta_i$ generate the same working-set sequence (see Proposition \ref{prop:unify}), the working set is the same at any iteration $\forall \theta \in \Theta_i$. Hence, if the processor state is the same at the start of an iteration, Assumption \ref{as:p-as} implies that the same sequence of processor instructions are executed in that iteration, and, therefore, that the processor state at the start of the next iteration will be the same $\forall \theta \in \Theta_i$. By induction, the processor instructions will be the same for \textit{any} iteration, i.e., exactly the same sequence of processor instructions will be executed altogether $\forall \theta \in \Theta_i$. Similar arguments hold with respect to cache hits/misses since, from Assumption \ref{as:p-as}, the memory access pattern is also completely determined by the working set.

    In conclusion, the execution time will be the same $\forall \theta \in \Theta_i$, given the same the starting processor state $q_0$. 
\end{\myproof}

\begin{corollary}
    \label{cor:wcet}
    Let the program $\mathcal{P}$ be a realization of an active-set method covered by the framework in \cite{arnstrom2022unifying}, and let $q$ denote the starting processor state; additionally, assume that $\mathcal{P}$ satisfies Assumption \ref{as:p-as}. Moreover, let $\{\Theta_i\}_{i=1}^N$ be all regions generated by the certification method presented in \cite{arnstrom2022unifying} and $\bar{\tau}_{\mathcal{P}}(q) \triangleq \max_{\theta \in \Theta_0} \tau_{\mathcal{P}}(\theta,q)$; then $\bar{\tau}_{\mathcal{P}}(q) = \max_{i \in \{1,\dots,N\}} \tau(\theta_i,q)$, for any collection of parameters $\{\theta_i\}_{i=1}^N$ such that $\theta_i \in \Theta_i$. 
\end{corollary}
\begin{\myproof}
    Follows from Lemma \ref{lem:etcover} and Theorem \ref{th:main}. 
\end{\myproof}

\begin{remark}
    \label{rem:more-than-wcet}
Note that we in addition to the worst-case for all $\theta \in \Theta_0$ know the \textit{exact} execution time given a starting processor state $q$, for any $\theta \in \Theta_0$. 
\end{remark}

To determine the WCET, one needs a way of determining the execution time $\tau_{\mathcal{P}}(\theta,q)$ for a given problem $\theta$ and processor state $q$. This can either be done by directly measuring the execution time on the hardware on which the active-set solver will be deployed, or by using a cycle-accurate simulator of the hardware; in the experiments in Section \ref{sec:result} we use the former. 

\subsection{Proposed WCET method}
\label{ssec:proposed-method}
Based on Corollary~\ref{cor:wcet}, we propose Algorithm \ref{alg:proposed} for determining the WCET of a program $\mathcal{P}$ that realizes Algorithm~\ref{alg:prot-as}. The grayed-out step, which will be motivated in Section \ref{ssec:reduction}, is optional and reduces the number of problems that have to be solved at Step \ref{step:measure}. 
The algorithm produces WCET estimates with the following properties:
\begin{theorem}
    Let $\tilde{\tau}_{\mathcal{P}}$ be the WCET estimate produced by Algorithm \ref{alg:proposed} and let $\bar{\tau}_{\mathcal{P}}$ be the true WCET defined in \eqref{eq:wcet}.
    If $q_0$ satisfy Assumption~\ref{as:start-q-determ}, then $\tilde{\tau}_{\mathcal{P}} = \bar{\tau}_{\mathcal{P}}$. 
    If, instead, $q_0$ satisfy Assumption~\ref{as:start-q-dominated}, then $\tilde{\tau}_{\mathcal{P}} \geq \bar{\tau}_{\mathcal{P}}$. 
\end{theorem}
\begin{\myproof}
    For the moment, disregard the pruning in Step~\ref{step:prune-subseq}.
    Given the properties of the inputs to the algorithm, the premises in Corollary \ref{cor:wcet} are satisfied. Hence, since the algorithm outputs $\tilde{\tau}_{\mathcal{P}} =  \max_{i\in\{1,\dots,N\}} \tau_{\mathcal{P}}(\theta_i, q_0)$, Corollary \ref{cor:wcet} implies that $\tilde{\tau}_{\mathcal{P}} = \bar{\tau}(q_0)$. If $q_0$ satisfies Assumption \ref{as:start-q-determ}, $\mathcal{P}$ will always start in $q_0$ and  we get $\bar{\tau}_{\mathcal{P}} = \bar{\tau}_{\mathcal{P}}(q_0)=\tilde{\tau}_{\mathcal{P}}$. If, on the other hand, $q_0$ satisfies Assumption \ref{as:start-q-dominated}, we get from Definition~\ref{def:dom} that $\tilde{\tau}_{\mathcal{P}} = \bar{\tau}_{\mathcal{P}}(q_0) \geq \bar{\tau}_{\mathcal{P}}$.

    The pruning in Step \ref{step:prune-subseq} will result in fewer execution of $\mathcal{P}$ in Step \ref{step:measure}. That omitting such executions does not affect the WCET estimate follows from Lemma \ref{lem:subseq}, presented in the upcoming Section \ref{ssec:reduction}.
\end{\myproof}
\begin{algorithm}
  \caption{Proposed WCET method}
  \label{alg:proposed}
  \begin{algorithmic}[1]
      \Require%
      {\begin{itemize}
              \item Parametric problem of the form \eqref{eq:opt} 
              \item Set of possible problem instances $\Theta_0$ 
              \item Program $\mathcal{P}$ satisfying Assumption \ref{as:p-as}
              \item Processor state $q_0$ satisfying Assumption \ref{as:start-q-determ} or \ref{as:start-q-dominated}
      \end{itemize}} 
      \Ensure WCET estimate $\tilde{\tau}_{\mathcal{P}}$ for $\mathcal{P}$
      \State $\{(\bar{k}_i, \mathbb{W}_i, \Theta_i)\}_{i=1}^N \leftarrow$ Use \cite{arnstrom2022unifying} on \eqref{eq:opt} for Alg.~\ref{alg:prot-as} and $\Theta_0$. \label{step:cert} 
      \State  $\mathcal{I} \leftarrow \{1,\dots,N \}$; \quad $\tilde{\tau} \leftarrow -\infty$
      \State \textcolor{gray}{Remove all $i$ such that $\exists j\neq i,  \mathbb{W}_j \succ \mathbb{W}_i$ from $\mathcal{I}$}\label{step:prune-subseq}
      \For {$i \in \mathcal{I}$}
      \State $\theta_i \leftarrow$ select any $\theta_i \in \Theta_i$ \label{step:cheby}
      \State Set processor state of $\mathcal{P}$ to $q_0$
      \State $\tau_i \leftarrow$ measure time for executing $\mathcal{P}$ with $\theta=\theta_i$ \label{step:measure}
      \State $\tilde{\tau} \leftarrow \max\{\tilde{\tau}, \tau_i\}$
      \EndFor
      \State \Return $\tilde{\tau}$
  \end{algorithmic}
\end{algorithm}

\subsection{Reducing the number of executed problems}
\label{ssec:reduction}

As Remark~\ref{rem:more-than-wcet} highlights, considering all regions produced by \cite{arnstrom2022unifying} gives more information than just the WCET: it gives the number of executed processor cycles \textit{for any} $\theta \in \Theta_0$. Next, we suggest a technique (used at Step \ref{step:prune-subseq} of Algorithm \ref{alg:proposed}) to reduce the set of problems that have to be solved on the hardware, without losing any information about the WCET. 

We do this by exploiting that some working-set sequences produced by the method in \cite{arnstrom2022unifying} are \textit{subsequences} of other produced working-set sequences.
\begin{definition}[Subsequence]
    A sequence of working sets $\tilde{\mathbb{W}} = \{\tilde{\mathcal{W}}_i\}_{i=1}^{|\tilde{\mathbb{W}}|}$ is a \textit{subsequence} of the sequence ${\mathbb{W} = \{{\mathcal{W}}_i\}_{i=1}^{|\mathbb{W}|}}$, denoted  $\tilde{\mathbb{W}} \prec \mathbb{W}$, if $|\mathbb{W}| > |\tilde{\mathbb{W}}|$ and $\mathcal{W}_i = \tilde{\mathcal{W}}_i$ for $i=1,\dots |\tilde{\mathbb{W}}|$, where $|\cdot|$ extracts the length of a sequence.
\end{definition}

From Assumption \ref{as:p-as} we have the following result (that motivates Step \ref{step:prune-subseq} in Algorithm \ref{alg:proposed}):
\begin{lemma}
    \label{lem:subseq}
    Let $\{\Theta_i\}^N_{i=1}$ be the regions produced by \cite{arnstrom2022unifying} and let $\mathbb{W}_i$ denote the working-set sequence generated by all $\theta \in \Theta_i$. If Assumption \ref{as:p-as} holds and $\mathbb{W}_j \prec \mathbb{W}_i$, then $\bar{\tau}(\theta_i,q) \geq \bar{\tau}(\theta_j,q)$, for all $\theta_i \in \Theta_i$ and $\theta_j \in \Theta_j$. 
\end{lemma}
\begin{\myproof}
    From Assumption~\ref{as:p-as}, $\mathbb{W}_i$ and $\mathbb{W}_j$ determines the sequence of executed processor state for any $\theta \in \Theta_i$  or $\theta \in \Theta_j$, respectively. Since $\mathbb{W}_j \prec \mathbb{W}_i$, the same processor instructions will be executed up until iteration $k = |\mathbb{W}_j|$ for all $\theta \in \Theta_i \cup \Theta_j$,  In addition, since $|\mathbb{W}_i| > |\mathbb{W}_j|$, additional processor instructions will be executed for $\theta \in \Theta_i$; consequently,
$\bar{\tau}(\theta_i,q) \geq \bar{\tau}(\theta_j,q)$ for all $\theta_i \in \Theta_i$ and $\theta_j \in \Theta_j$.
\end{\myproof}

\section{Numerical Experiments}
\label{sec:result}
In this section we perform experiments to validate the proposed WCET method (Algorithm \ref{alg:proposed}). Our goals with the experiments are: ($i$) to validate the correctness of our method (Section~\ref{ssec:res-valid}); ($ii$) to show that our method accounts for implementation-specific choices such as compilation flags (Section~\ref{ssec:res-flags}); ($iii$) to compare the method with classical static and measurement-based WCET methods (Section~\ref{ssec:res-comp}); ($iv$) to show how the method scales and how effective the pruning from Lemma \ref{lem:subseq} is (Section~\ref{ssec:res-scalability}).

We consider the classical MPC problem of stabilizing an inverted pendulum on a cart. The objective is to stabilize the system at a given reference point (angle and displacement), while accounting for actuator constraints.
The MPC uses a horizon of $10$ time-steps, resulting in a problem of the form~\eqref{eq:opt} with the dimensions $n_x = 10$, $n_{\theta} = 8$, and $m=20$; see, e.g., Section VI in \cite{cimini2017certqp} for more details regarding the MPC problem. 

The active-set solver in \cite{arnstrom2022daqp} (DAQP), which implements Algorithm \ref{alg:prot-as}, is used to solve the resulting QPs. For all experiments, DAQP was initialized with an empty working set ($\mathcal{W}_0 =\emptyset$). The target hardware is an STM32F411 MCU, which uses an ARM Cortex-M4 core. In the experiments we use the default clock frequency of 84 GHz. The MCU have 512 kB of flash memory and 128kB of SRAM. The core itself has no caches, but the MCU has an instruction prefetch cache connected to its flash memory. The GNU Compiler Collection (GCC) was used for compilation. 

To fulfill Assumption~\ref{as:start-q-determ}, the instruction prefetch cache of the MCU is flushed before solving a problem to ensure a deterministic initial processor state. 
To fulfill Assumption \ref{as:p-as}, a C-implementation of DAQP was adapted based on the implementation guidelines in Appendix~\ref{ap:assumption-details}, for example by using the order-independent minimization outlined in Algorithm~4.

The number of processor cycles is recorded with an on-board timer with a resolution of 3 cycles (a higher resolution leads to overflow). Given the number of cycles, the corresponding execution time can be obtained by dividing this number with the clock frequency (84 GHz). In this article we just report the number of cycles for simplicity. 

\subsection{Empirical validation of Algorithm \ref{alg:proposed}}
\label{ssec:res-valid}
The maximum number of processor cycles determined by Algorithm \ref{alg:proposed}, for any $\theta \in \Theta_0 \subseteq \mathbb{R}^8$, and with the compiler flag \texttt{-O0}, was 136809 cycles.
The time for running the certification method from \cite{arnstrom2022unifying} was 14 seconds, which generated 89149 problem instances to execute on the MCU. To select a parameter in each region, the Chebyshev center for each region was computed (the corresponding times for computing these are included in the 14 seconds taken by the certification method.) Solving all 89149 QP instances on the MCU took 627 seconds (with the compilation flag \texttt{-O0}.)

To validate the result, $10^6$ samples were randomly selected from the parameter space;  for each sample, the corresponding QP  was solved on the hardware and the executed number of processor cycles was measured. This was then verified to be equal to the expected number of cycles from solving the archetypal problem for the region in which the sample resides, which supports the correctness of Algorithm \ref{alg:proposed}. 

\subsection{Handling of implementation-specific choices}
\label{ssec:res-flags}
\begin{figure}
  \centering
  \includegraphics[width=0.5\textwidth]{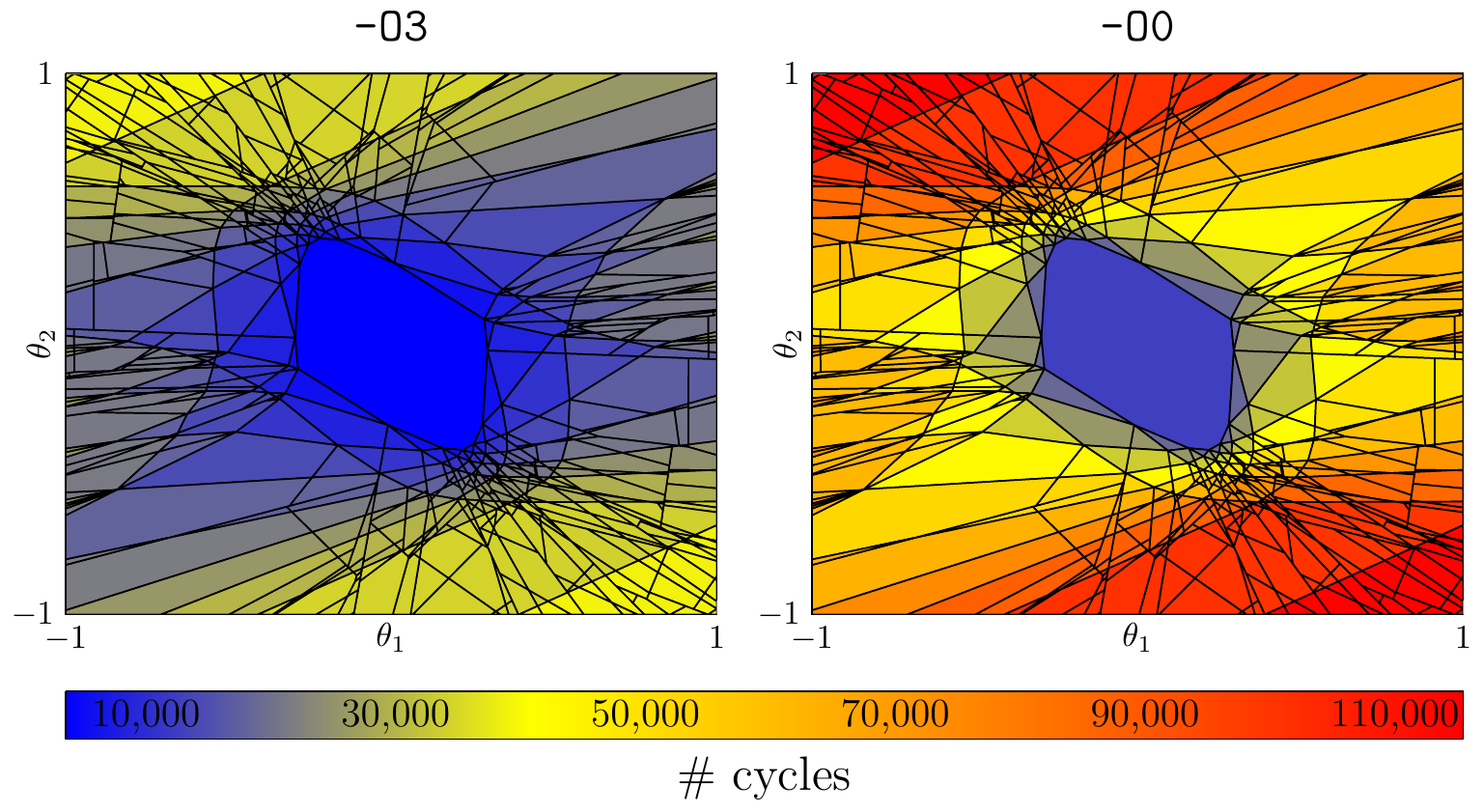}
  \caption{Number of cycles required by the program for different problem instances and compilation flags, certified by the proposed method.}
  \label{fig:invpend-cert}
\end{figure}

Figure~\ref{fig:invpend-cert} illustrates a two dimensional slice of the result that the proposed method produces.  Specifically, it shows a two-dimensional slice of the parameter space ($\theta_i = 0$ for $i=3,\dots,8$), where each point represents a QP that might need to be solved online. Parameters in the same region generate the same working-set sequence to determine a solution, and therefore, from Theorem~\ref{th:main}, execute the same number of processor cycles. 

To exemplify that our WCET method can account for implementation-specific choices, Figure~\ref{fig:invpend-cert} also depicts resulting partitions for two different compilation flags passed to GCC; the flag \texttt{-O3} means ``full code optimization'', and \texttt{-O0} means ``no code optimization''. As is to be expected, significantly fewer processor cycles are required when compiler optimization is enabled.  

\begin{figure}
  \centering
  \includegraphics[width=0.2\textwidth]{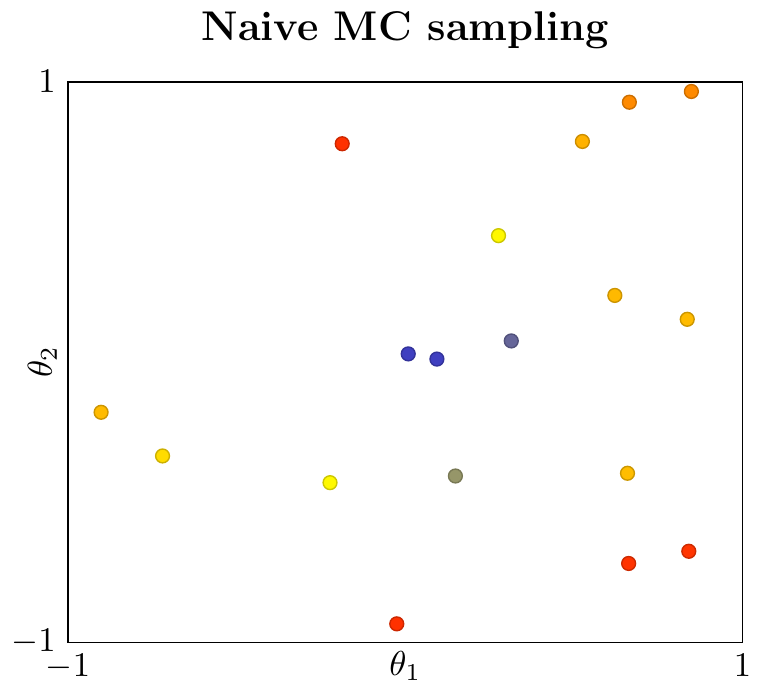}
  \caption{Number of cycles required by the program (compiled with \texttt{-O0}), for $89768^{\frac{2}{8}}\approx 18$ randomly selected QP problems in a two-dimensional slice of $\Theta_0$. This corresponds to the same number of QP problems (projected from 8 to 2 dimensions) solved by the proposed method.}
  \label{fig:invpend-mc}
\end{figure}

\subsection{Comparison with classical WCET methods}
\label{ssec:res-comp}
For the resulting program, we also used the state-of-the art (proprietary) static WCET analyzer  
aiT \cite{ferdinand2004ait}. The analysis took 215 seconds and gave a worst-case bound of ${5.7\cdot 10^6}$ cycles, which is 40 times higher than the actual WCET.
The WCET bound generated by aiT could be improved somewhat by specifying more detailed flow facts \cite{gustafsson2006automatic}, but this often requires extensive work by the user, and, still, will never reach the tight WCET provided by the proposed method. Note, however, that aiT can be applied to generic programs while ours is specialized for programs that realize Algorithm \ref{alg:prot-as}.

Figure \ref{fig:invpend-mc} exemplify how a classical measurement-based approach would sample the parameter space if as many measurements (i.e., solved QPs) are allowed as in the proposed method. Clearly, the coverage of possible QP problems is not sufficient for a reliable WCET estimate; but the same number of QP problems suffice in the proposed method because they are conscientiously selected based on Theorem~\ref{th:main}. Moreover, even if more samples were allowed to be taken in a classical measurement-based approach, the resulting WCET can never be guaranteed (in contrast to the proposed method). 

\subsection{Scalability and effectiveness of pruning}
\label{ssec:res-scalability}
In Table~\ref{tab:invpend} we report results from applying Algorithm \ref{alg:proposed} on problems of different dimensions by using different horizons in the MPC. 
For horizon $N$, the resulting dimensions of the optimization problems are $n=N$, $m=2N$ and $n_{\theta} = 8$. The time taken to determine archetypal problems (i.e., the time for Step \ref{step:cert} and \ref{step:cheby}) is denoted $t_{\text{cert}}$. The time taken to solve problems on the hardware (i.e., the time for Step \ref{step:measure}) is denoted $t_{\text{sim}}$. We make use of Lemma \ref{lem:subseq}, so only a subset of archetypal problems (also reported in Table \ref{tab:invpend}) need to be solved on the hardware.
The result reported in Table \ref{tab:invpend} can be used for an exact trade-off between computational effort vs horizon length, which can be used to select the most suitable horizon length for the given application and hardware at hand. 
\begin{table}
    \caption{Algorithm \ref{alg:proposed} applied to the multi-parameter quadratic programs originating from the MPC of an inverted pendulum with varying horizon $N$. The compiler flag $\texttt{-O0}$ was used.}
    \label{tab:invpend}
    \centering

    \begin{filecontents*}{data/invpend.dat}
N   nth     nthprune  tcert  tsim tsimprune maxcycle  meancycle 
6   2003    900       0.1   -     6       48543     37867 
8   15669   8236      2      110  58        85584     67579 
10  89149   52637     14     627  370       136809    108246
12  407677  261341    109    -    1919      228870    162152
14  1543888 1020459   616    -    9094      322566    227506
\end{filecontents*}
\pgfplotstabletypeset[
    columns={N,nth, nthprune, tcert, tsimprune, maxcycle},
    columns/N/.style={
        column name={\centering $N$},
        column type=c|,
    },
    columns/nth/.style={
        column name={Nominal},
        column type=r,
        fixed,
        fixed zerofill,
        precision=0,
    },
    columns/nthprune/.style={
        column name={Lemma \ref{lem:subseq}},
        column type=r|,
        fixed,
        fixed zerofill,
        precision=0,
    },
    columns/tcert/.style={
        column name={$t_{\text{CERT}}$},
        column type=r,
    },
    columns/tsimprune/.style={
        column name={$t_{\text{SIM}}$},
        column type=r|,
    },
    columns/maxcycle/.style={
        column name={worst-case},
        column type=r,
        fixed,
        fixed zerofill,
        precision=0,
    },
    every head row/.style={
        before row={
            \toprule
            \multicolumn{1}{c}{}
    &\multicolumn{2}{c}{\# archetypal problems}
    &\multicolumn{2}{c}{Time [s]}
    &\multicolumn{1}{c}{\# cycles}\\
},
after row=\midrule
},
every last row/.style={after row=\bottomrule},
]
{data/invpend.dat} 
\end{table}

\section{Conclusion}
We have proposed a method that determines the worst-case execution time (WCET) for linear model predictive control (MPC). The method leverages the complexity certification framework in \cite{arnstrom2022unifying} to generate a finite set of ``archetypal'' optimization problems.
    By solving just these archetypal problems on the hardware on which the MPC is to be deployed, and by recording the execution times, we obtain the \textit{exact} WCET.
    The proposed method was validated by considering MPC of an inverted pendulum on a cart, which illustrated advantages compared with classical WCET methods: ($i$) in contrast to classical static methods, our method gives the \textit{exact} WCET; ($ii$) in contrast to classical measurement-based methods, our method guarantees a correct WCET estimate and require fewer measurements on the hardware.

\bibliographystyle{IEEEtran}
\linespread{1.0}\selectfont

\bibliography{tex/lib.bib}

\begin{thebibliography}{10}
\providecommand{\url}[1]{#1}
\csname url@samestyle\endcsname
\providecommand{\newblock}{\relax}
\providecommand{\bibinfo}[2]{#2}
\providecommand{\BIBentrySTDinterwordspacing}{\spaceskip=0pt\relax}
\providecommand{\BIBentryALTinterwordstretchfactor}{4}
\providecommand{\BIBentryALTinterwordspacing}{\spaceskip=\fontdimen2\font plus
\BIBentryALTinterwordstretchfactor\fontdimen3\font minus
  \fontdimen4\font\relax}
\providecommand{\BIBforeignlanguage}[2]{{%
\expandafter\ifx\csname l@#1\endcsname\relax
\typeout{** WARNING: IEEEtran.bst: No hyphenation pattern has been}%
\typeout{** loaded for the language `#1'. Using the pattern for}%
\typeout{** the default language instead.}%
\else
\language=\csname l@#1\endcsname
\fi
#2}}
\providecommand{\BIBdecl}{\relax}
\BIBdecl

\bibitem{rawlings2017model}
J.~Rawlings, D.~Mayne, and M.~Diehl, \emph{Model Predictive Control: Theory,
  Computation, and Design}.\hskip 1em plus 0.5em minus 0.4em\relax Nob Hill
  Publishing, 2017, vol.~2.

\bibitem{di2018real}
S.~Di~Cairano and I.~V. Kolmanovsky, ``Real-time optimization and model
  predictive control for aerospace and automotive applications,'' in \emph{2018
  American Control Conference (ACC)}, 2018, pp. 2392--2409.

\bibitem{johansen2014dmpc}
T.~A. Johansen, ``Toward dependable embedded model predictive control,''
  \emph{IEEE Systems Journal}, vol.~11, no.~2, pp. 1208--1219, 2014.

\bibitem{bemporad2002explicit}
A.~Bemporad, M.~Morari, V.~Dua, and E.~N. Pistikopoulos, ``The explicit linear
  quadratic regulator for constrained systems,'' \emph{Automatica}, vol.~38,
  no.~1, pp. 3--20, 2002.

\bibitem{cimini2017certqp}
G.~Cimini and A.~Bemporad, ``Exact complexity certification of active-set
  methods for quadratic programming,'' \emph{IEEE Transactions on Automatic
  Control}, vol.~62, pp. 6094--6109, 2017.

\bibitem{domahidi2012efficient}
A.~Domahidi, A.~U. Zgraggen, M.~N. Zeilinger, M.~Morari, and C.~N. Jones,
  ``Efficient interior point methods for multistage problems arising in
  receding horizon control,'' in \emph{2012 IEEE 51st IEEE conference on
  decision and control (CDC)}.\hskip 1em plus 0.5em minus 0.4em\relax IEEE,
  2012, pp. 668--674.

\bibitem{ferreau2014qpoases}
H.~J. Ferreau, C.~Kirches, A.~Potschka, H.~G. Bock, and M.~Diehl, ``{qpOASES}:
  A parametric active-set algorithm for quadratic programming,'' \emph{Math.
  Program. Comput.}, vol.~6, pp. 327--363, 2014.

\bibitem{frison2020hpipm}
G.~Frison and M.~Diehl, ``{HPIPM}: a high-performance quadratic programming
  framework for model predictive control,'' \emph{IFAC-PapersOnLine}, vol.~53,
  no.~2, pp. 6563--6569, 2020.

\bibitem{patrinos2014accelerated}
P.~Patrinos and A.~Bemporad, ``An accelerated dual gradient-projection
  algorithm for embedded linear model predictive control,'' \emph{IEEE
  Transactions on Automatic Control}, vol.~59, no.~1, pp. 18--33, 2014.

\bibitem{arnstrom2022daqp}
D.~Arnström, A.~Bemporad, and D.~Axehill, ``A dual active-set solver for
  embedded quadratic programming using recursive {LDL}$^{T}$ updates,''
  \emph{IEEE Trans. Autom. Control}, vol.~67, no.~8, pp. 4362--4369, 2022.

\bibitem{nocedal}
J.~Nocedal and S.~Wright, \emph{\BIBforeignlanguage{en}{Numerical
  Optimization}}.\hskip 1em plus 0.5em minus 0.4em\relax Springer Science \&
  Business Media, 2006.

\bibitem{fletcher1971general}
R.~Fletcher, ``A general quadratic programming algorithm,'' \emph{IMA Journal
  of Applied Mathematics}, vol.~7, no.~1, pp. 76--91, 1971.

\bibitem{goldfarb1983dual}
D.~Goldfarb and A.~Idnani, ``A numerically stable dual method for solving
  strictly convex quadratic programs,'' \emph{Mathematical Programming},
  vol.~27, pp. 1--33, 9 1983.

\bibitem{klee1972good}
V.~Klee and G.~J. Minty, ``How good is the simplex algorithm,''
  \emph{Inequalities}, vol.~3, no.~3, pp. 159--175, 1972.

\bibitem{spielman2004smoothed}
D.~A. Spielman and S.-H. Teng, ``Smoothed analysis of algorithms: Why the
  simplex algorithm usually takes polynomial time,'' \emph{Journal of the ACM
  (JACM)}, vol.~51, no.~3, pp. 385--463, 2004.

\bibitem{arnstrom2022unifying}
D.~Arnström and D.~Axehill, ``A unifying complexity certification framework
  for active-set methods for convex quadratic programming,'' \emph{IEEE Trans.
  Autom. Control}, vol.~67, no.~6, pp. 2758--2770, 2022.

\bibitem{wilhelm2008worst}
R.~Wilhelm, J.~Engblom, A.~Ermedahl, N.~Holsti, S.~Thesing, D.~Whalley,
  G.~Bernat, C.~Ferdinand, R.~Heckmann, T.~Mitra \emph{et~al.}, ``The
  worst-case execution-time problem—overview of methods and survey of
  tools,'' \emph{ACM Transactions on Embedded Computing Systems (TECS)},
  vol.~7, no.~3, pp. 1--53, 2008.

\bibitem{ferdinand2004ait}
C.~Ferdinand and R.~Heckmann, ``{aiT}: Worst-case execution time prediction by
  static program analysis,'' in \emph{Building the Information Society}.\hskip
  1em plus 0.5em minus 0.4em\relax Springer, 2004, pp. 377--383.

\bibitem{ballabriga2010otawa}
C.~Ballabriga, H.~Cass{\'e}, C.~Rochange, and P.~Sainrat, ``Otawa: An open
  toolbox for adaptive {WCET} analysis,'' in \emph{IFIP International Workshop
  on Software Technolgies for Embedded and Ubiquitous Systems}.\hskip 1em plus
  0.5em minus 0.4em\relax Springer, 2010, pp. 35--46.

\bibitem{broman2017brief}
D.~Broman, ``A brief overview of the {KTA WCET} tool,'' \emph{arXiv preprint
  arXiv:1712.05264}, 2017.

\bibitem{davis2019survey}
R.~I. Davis and L.~Cucu-Grosjean, ``A survey of probabilistic timing analysis
  techniques for real-time systems,'' \emph{LITES: Leibniz Transactions on
  Embedded Systems}, pp. 1--60, 2019.

\bibitem{natarajan2019weaklyhard}
S.~Natarajan, M.~Nasri, D.~Broman, B.~B. Brandenburg, and G.~Nelissen, ``From
  code to weakly hard constraints: A pragmatic end-to-end toolchain for {Timed
  C},'' in \emph{2019 IEEE Real-Time Systems Symposium (RTSS)}, 2019, pp.
  167--180.

\bibitem{axer2014building}
P.~Axer, R.~Ernst, H.~Falk, A.~Girault, D.~Grund, N.~Guan, B.~Jonsson,
  P.~Marwedel, J.~Reineke, C.~Rochange \emph{et~al.}, ``Building timing
  predictable embedded systems,'' \emph{ACM Transactions on Embedded Computing
  Systems (TECS)}, vol.~13, no.~4, pp. 1--37, 2014.

\bibitem{zimmer2014flexpret}
M.~Zimmer, D.~Broman, C.~Shaver, and E.~A. Lee, ``{FlexPRET}: A processor
  platform for mixed-criticality systems,'' in \emph{2014 IEEE 19th Real-Time
  and Embedded Technology and Applications Symposium (RTAS)}, 2014, pp.
  101--110.

\bibitem{cousot1977abstract}
P.~Cousot and R.~Cousot, ``Abstract interpretation: a unified lattice model for
  static analysis of programs by construction or approximation of fixpoints,''
  in \emph{Proceedings of the 4th ACM SIGACT-SIGPLAN symposium on Principles of
  programming languages}, 1977, pp. 238--252.

\bibitem{ferdinand1999efficient}
C.~Ferdinand and R.~Wilhelm, ``Efficient and precise cache behavior prediction
  for real-time systems,'' \emph{Real-time systems}, vol.~17, pp. 131--181,
  1999.

\bibitem{touzeau2019cache}
V.~Touzeau, C.~Ma\"{\i}za, D.~Monniaux, and J.~Reineke, ``Fast and exact
  analysis for lru caches,'' \emph{Proc. ACM Program. Lang.}, vol.~3, no. POPL,
  jan 2019.

\bibitem{reineke2006definition}
J.~Reineke, B.~Wachter, S.~Thesing, R.~Wilhelm, I.~Polian, J.~Eisinger, and
  B.~Becker, ``A definition and classification of timing anomalies,'' in
  \emph{6th International Workshop on Worst-Case Execution Time Analysis
  (WCET'06)}.\hskip 1em plus 0.5em minus 0.4em\relax Schloss
  Dagstuhl-Leibniz-Zentrum f{\"u}r Informatik, 2006.

\bibitem{arnstrom2021lic}
D.~Arnstr{\"o}m, ``On complexity certification of active-set {QP} methods with
  applications to linear {MPC},'' Licentiate Thesis, Link{\"o}ping University
  Electroinic Press, 2021.

\bibitem{wong2011active}
E.~Wong, \emph{Active-set methods for quadratic programming}.\hskip 1em plus
  0.5em minus 0.4em\relax University of California, San Diego, 2011.

\bibitem{gustafsson2006automatic}
J.~Gustafsson, A.~Ermedahl, C.~Sandberg, and B.~Lisper, ``Automatic derivation
  of loop bounds and infeasible paths for {WCET} analysis using abstract
  execution,'' in \emph{2006 27th IEEE International Real-Time Systems
  Symposium (RTSS'06)}.\hskip 1em plus 0.5em minus 0.4em\relax IEEE, 2006, pp.
  57--66.

\end{thebibliography}
 \appendix
 \subsection{Details on Assumption \ref{as:p-as}}
 \label{ap:assumption-details}
 To show that Assumption \ref{as:p-as} is practically sound, we expand upon how the active-set solver in \cite{arnstrom2022daqp} can be implemented to fulfill it. The main point is to show that the particular value of $\theta$ does not alter the executed arithmetic operations, nor the memory accesses, in an iteration. Even though we focus on the solver in \cite{arnstrom2022daqp}, similar arguments apply to other active-set methods covered by \cite{arnstrom2022unifying}, e.g., the ones in \cite{nocedal,fletcher1971general,goldfarb1983dual}. 
 \subsubsection{Solving linear system of equations (Step \ref{step:linsys} and \ref{step:linsys-sing})}
 \label{ap:linsys}
 The linear systems of equations solved in Step \ref{step:linsys} and \ref{step:linsys-sing} are of the form $[M]^T_{\mathcal{W}_k}[M]_{\mathcal{W}_k} \lambda^* = -[d(\theta)]_{\mathcal{W}_k}$, where $M$ is a given matrix, and $d$ is an affine function. Moreover, $[\cdot]_{\mathcal{W}_k}$ denotes extraction of the rows of a matrix/vector in the index set $\mathcal{W}_k$ (hence, $\mathcal{W}_k$ determines the dimension of the linear system of equations and which elements of $M$ and $d$ need to be accessed in memory).  Importantly, the parameter $\theta$ only enters in the right-hand-side of the equation. If this system is solved using a factorization-based solution method, the same arithmetic operation will be performed for all $\theta$ (different $\theta$ only yields different \textit{operands}, not different \textit{operations}) since the left-hand-side is parameter-independent; moreover, the memory accesses are also parameter-independent.
Note that this might, however, not be the case if instead an iterative method is used to solve the linear system of equations. 

\subsubsection{Affine transformation (Step~\ref{step:affine-transformation})}
The variable $\mu$ in an iteration is given by $\mu = [M]_{\overline{\mathcal{W}}} [M]_{\mathcal{W}} \lambda^*(\theta) + [d(\theta)]_{\overline{\mathcal{W}}}$, where $\lambda^*$ is computed as described above (and will be affine in $\theta$) and $\overline{\mathcal{W}}$ denotes the complement of $\mathcal{W}$, i.e., it contains \emph{inactive} constraints. Hence, the dimension of the matrix $[M]_{\overline{\mathcal{W}}} [M]_{\mathcal{W}}$ and the vector $[d(\theta)]_{\overline{\mathcal{W}}}$ only depends on $\mathcal{W}$, so the arithmetic operations for the affine transformation only depend on $\mathcal{W}$ (recall that $M$ and $d$ are given). Moreover, for a given $\mathcal{W}$, the same rows of $M$ and $d$ are indexed, resulting in the same memory accesses. 

\subsubsection{Selecting a constraint to add to $\mathcal{W}$ (Step \ref{step:add-constr})}
When selecting a constraint to add to $\mathcal{W}$ at Step \ref{step:add-constr}, any selection $i \notin \mathcal{W}$ such that $[\mu]_i < 0$ suffices. The most common selection rule is Dantzig's selection rule defined by ${i = \argmin_{j\notin \mathcal{W}} [\mu(\theta)]_j}$. Another popular rule is Bland's selection rule, which selects the lexicographically minimal $i$, i.e., $i = \min_{j\notin \mathcal{W}: [\mathcal{\mu}]_j<0} j$. 

The properties of the certification method in \cite{arnstrom2022unifying} generate regions such that all parameters in a particular region lead to the same selection $i$. The interdependent ordering among the elements might, however, vary for different values of $\theta$ in the same region. Hence, if care is not taken when implementing the selection rule, the execution time for an archetypal problem instance might not coincide with the execution time for other problem instances residing in the same region. Such discrepancies can be avoided in two ways: first, the partitioning in \cite{arnstrom2022unifying} can be extended to produce regions in which the ordering among the elements in $\mu(\theta)$ is the same. This is a straightforward extension, but increases the required computations in the certification.  The second way, which we use in the experiments, is to make the computational complexity of $\argmin_j$ independent of the actual values in the vector. A comparison of an order-dependent and an order-independent implementation of $\argmin$ is given in Algorithm 3 and 4, respectively. 
\begin{algorithm}[H]
    \small
    \label{alg:order-dep}
    \caption{Order-dependent algorithm for finding the minimizing element and index $v,i$ for a vector $\mu$.}
  \begin{algorithmic}[1]
      \State $v \leftarrow \infty$;\:\: $i \leftarrow -1$
      \For {$j \in \{1,\dots, N$\} \textbf{if} $[\mu(\theta)]_j < v$ \textbf{then}}
      \State $v \leftarrow [\mu(\theta)]_j$; \quad $i \leftarrow j$
      \EndFor
  \end{algorithmic}
  \normalsize
\end{algorithm}
\begin{algorithm}[H]
    \small
    \label{alg:order-indep}
    \caption{Order-independent algorithm for finding the minimizing element and index $v,i$ for a vector $\mu$. A true/false expression should be interpreted as 1/0, respectively. 
    }
  \begin{algorithmic}[1]
      \State $v \leftarrow \infty$;\:\: $i \leftarrow -1$
      \For {$j \in \{1,\dots, N$\}}
      \State $v \leftarrow ([\mu(\theta)]_j < v) \cdot [\mu(\theta)]_j + ([\mu(\theta)]_j \geq v) \cdot v$
      \State $i\: \leftarrow ([\mu(\theta)]_j < v) \cdot j  + ([\mu(\theta)]_j \geq v) \cdot i$
      \EndFor
  \end{algorithmic}
  \normalsize
\end{algorithm}

\subsubsection{Selecting a constraint to remove from $\mathcal{W}$ (Step \ref{step:rm-constr})}
The removal of a constraint $i$ from $\mathcal{W}$, performed at Step \ref{step:rm-constr} of Algorithm \ref{alg:prot-as}, is done based on the ratio test
${i = \argmin_{j\in \mathcal{W}: [\lambda^*(\theta)]_j < 0} \frac{[\lambda^*(\theta)]_j}{[\lambda^*(\theta)-\lambda(\theta)]_j}}$.
Similar to the case of adding a constraint to $\mathcal{W}$ described above, the certification method ensures that the same $i$ is selected for all parameters in a particular region, but not that the interdependent ordering among the elements of $\lambda^*$ are the same. Hence, similar care has to be taken when implementing it.

\end{document}